# Fine-Grained Authorization for Job and Resource Management Using Akenti and the Globus Toolkit®

M. R. Thompson, A. Essiari
*LBNL, Berkeley, CA 94705*

K. Keahey, V, Welch, S. Lang
*ANL, Argonne IL 60439*

B. Liu
*University of Houston, Houston, TX 77204*

As the Grid paradigm is adopted as a standard way of sharing remote resources across organizational domains, the need for fine-grained access control to these resources increases. This paper presents an authorization solution for job submission and control, developed as part of the National Fusion Collaboratory, that uses the Globus Toolkit 2 and the Akenti authorization service in order to perform fine-grained authorization of job and resource management requests in a Grid environment. At job startup, it allows the system to evaluate a user's Resource Specification Language request against authorization policies on resource usage (determining how many CPUs or memory a user can use on a given resource or which executables the user can run). Furthermore, based on authorization policies, it allows other virtual organization members to manage the user's job.

## 1. INTRODUCTION

Users from different organizations who are geographically dispersed but are working together to solve a common problem, or related problems in a common domain, typically organize themselves into virtual organizations (*VOs*) [5]. The VO defines who its members are and (possibly) assigns roles or attributes to the members. The VO also arranges with the owners of various resources for VO member access. The resources may consist of compute platforms, storage elements, scientific instruments, data or services.

The National Fusion Collaboratory (*NFC*) [8] is an example of such a VO. The NFC is building a FusionGrid to provide computational and data services to its members. Because the Globus Toolkit (*GT2*) [6] is so widely used as Grid middleware, the NFC has chosen to use GT2 for remote job submission and secure access to its common data servers.

While object-oriented distributed programming frameworks such as Legion [4] and CORBA provide very fine-grained access-control at the level of object methods, GT2 provides a coarse-grained "admission control" facility and leaves fine-grained access control up to the resource provider. This simple approach is entirely acceptable for the initial stages of a Grid, when there is a limited set of potential users who negotiate access directly with the resource providers, but it does not scale to large numbers of resource hosts and users.

Hence, GT2 access control mechanisms must be extended to meet the FusionGrid's security needs. The solution we present here is to integrate the Akenti authorization service [9] with the Globus Toolkit.

Section 2 of this paper describes typical usage scenarios for VO Grid use. Section 3 is a brief overview of how authorization is currently handled in GT2. Section 4 introduces the Akenti authorization service. Section 5 describes our integration of the Globus Toolkit job manager and Akenti authorization and how this model can be extended to other authorization decision functions. Section 6 presents our conclusions and outlines future work.

## 2. USAGE SCENARIOS AND REQUIREMENTS

Many different resource-sharing scenarios exist in a Grid envirnoment. The shared resources may be basic compute resources (e.g., compute cycles and storage elements); sophisticated computer-controlled instruments; data elements such as files and information in databases; or services provided by specialized application programs. Individual resource providers may want detailed control over user access, or they may want to delegate most of the control to the VO. Multiple independent entities, called *stakeholders*, may be entitled to some control over a resource. For example, application code may be provided by one person or organization and run on a computer provided by an independent organization.

The use case that we are addressing in the NFC is that of an application service provider [12] where both the code and the compute resources are owned by the same entity. Selected hosts within the NFC allow remote users to execute specific codes. The FusionGrid has several sites that are providing access to a limited number of application codes. Thus, the sites want to restrict which executables may be run. Since these are computationally intensive codes that may take a long time to complete, the ability to query and control a job is important. Thus jobs become dynamic resources that need access control. The NFC wants to allow some of its users access to development versions of the code and tools in addition to the service codes. It may also want to allow some users a higher quality of service.

In order to support fine-grained access, the access control decision function (*ADF*) must be able to base its access decisions on policy written in a moderately expressive policy language. Such a language must be easy for stakeholders to understand and must be extensible to allow for many types of resources and conditions.

In summary, the challenging access control requirements that we address are as follows:
- Providing flexible policy-driven access control
- Federating policy from several independent sources
- Allowing long-running jobs to be treated as objects whose management is subject to access control decisions
- Integrating with the current GT2 job submission mechanism with a minimum disturbance for the client or the service provider

## 3. AUTHORIZATION IN THE GLOBUS TOOLKIT

We assume the following model for job submission and control. An interaction is initiated by a user submitting a request to start a job, including the job description, accompanied by the user's Grid credentials, in the form of an X.509 certificate [7]. In the current case this is just an identity certificate and asserts no other attributes about the user. This request is then evaluated by an access control decision function (*ADF*) which may be called from several different access control enforcement functions (*AEF*s) located in the resource management modules. If the request is authorized, it is started under a local credential (i.e., userid).

During the job execution, a VO user may submit management requests composed of a management action (e.g., request information, suspend or resume a job, cancel a job). The resource manager may decide to perform the action or to pass it on to the locally executing job.

In order to perform these transactions, the Globus Resource Acquisition and Management (*GRAM*) [2] system is used. GRAM has two major software components: the gatekeeper and the job manager. The gatekeeper is responsible for translating Grid credentials to local credentials (e.g. mapping the user to a local account based on their Grid credentials) and creating a job manager instance to handle the specific job invocation request. The job manager is a Grid service which instantiates and then provides for the ability to manage a job. Figure 1 shows the interaction of these elements; in this section we explain their roles and limitations.

### 3.1. Gatekeeper

The GRAM gatekeeper is responsible for authenticating the requesting Grid user, authorizing a job invocation request, and determining the account in which the job should be run. Authentication, done using the Globus Toolkit's Grid Security Infrastructure (GSI) [1], verifies the validity of the presented Grid credentials, the user's possession of those credentials, and the user's Grid identity as indicated by those credentials. Authorization is based on the user's Grid identity, the site's trust policy, and the site grid-mapfile, which maps each allowed Grid identity to a local userid.

The gatekeeper then starts a job manager instance, executing with the user's local credential. This mode of operation requires the user to have an account on the resource and implements fine-grained access enforcement by privileges of the account.

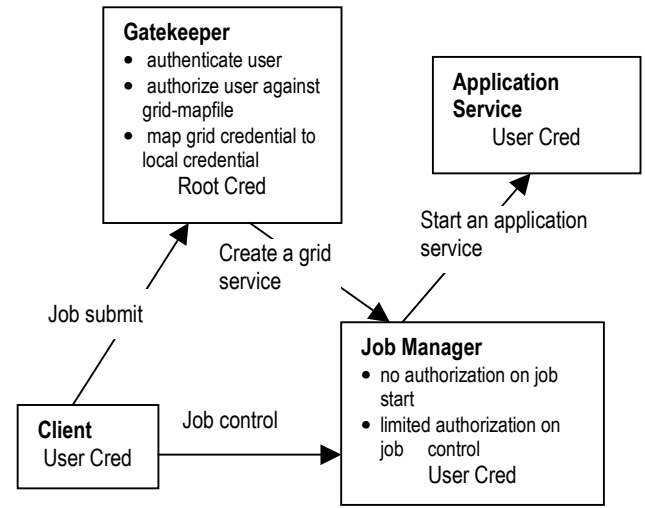

Figure 1 Interaction of the main components of GRAM

### 3.2. Job Manager

The GRAM job manager parses the user's request, including the job description, and calls the resource's job control system (e.g., exec, LSF, PBS) to initiate the user's job. During the job execution the job manager monitors its progress and handles job control requests (e.g., suspend, stop, query) from the user. Since the job manager instance is run under the user's local credential, as defined by the user's account, the operating system, and local job control system are able to enforce local policy on the job manager and user job by the policy tied to that account.

The job manager does no authorization on job startup because the gatekeeper has already done so. Once the job has been started, however, the job manager accepts, authenticates, and authorizes management requests on the job.

In GT2, the authorization policy on these management requests is static and simple: the Grid identity of the user making the request must match the Grid identity of the user who initiated the job.

## 4. AKENTI AUTHORIZATION SERVICE

As noted in Section 1, the authorization provided by GT2 is coarse grained. Because of the large user community, the NFC needed to add fine-grained authorization for job execution and management. Rather than writing an authorization function from scratch, the NFC decided to use the Akenti authorization service [10]. Akenti is an established authorization service designed to make access decisions for distributed resources controlled by multiple stakeholders. Akenti assumes that all the parties involved in authorization have X.509 certificates that can be used for identification and authentication. Authorization policy for a resource is represented as a set of (possibly) distributed certificates digitally signed by unrelated stakeholders from different domains. These policy certificates are independently created by authorized stakeholders. When an authorization decision needs to be made, the Akenti policy engine gathers all the relevant certificates for the user and the resource, verifies them, and determines the user's rights with respect to the resource.

### 4.1. Authorization Model

The Akenti model consists of *resources* that are being accessed via a *resource gateway* (the AEF) by *clients*. These clients connect to the resource gateway using the TLS [3] handshake protocol, or something equivalent, to present authenticated X.509 certificates. The *stakeholders* for the resources express *access constraints* on the resources as a set of *signed certificates*, a few of which are self-signed and must be stored on a known secure host (probably the resource gateway machine), but most of which can be stored remotely. These certificates express the attributes a user must have in order to get specific rights to a resource, identify the stakeholders who are trusted to create use-condition statements, and determines the attribute authorities who can attest to a user's attributes. At the time of the resource access, the resource gatekeeper (AEF) asks a trusted Akenti server (ADF) what access the user has to the resource. The Akenti server finds all the relevant certificates, verifies that each one is signed by an acceptable issuer, evaluates them, and returns the allowed access.

Several models for authorization systems have been proposed. One is the *pull model,* in which the user presents only his authenticated identity to the gatekeeper, who finds (pulls) the policy information for the resource and evaluates the user's access. Another model is the *push model*, in which the user presents one or more tokens or assertions that grant the holder specific rights to the resource. In this model, the gatekeeper must verify that the user has the rights to use the tokens and then must interpret the rights that have been presented.

In the application shown in Figure 2, the pull model is used in order to allow applications to transparently use Akenti authorization over standard GSI/TLS connections that transport and verify X.509 certificates. Akenti can also be used in a push model because it returns its authorization decision as a signed capability certificate containing the subject's distinguished name (DN), public key, the certification authority (*CA*) that signed for this name, the name of the resource, and the subject's rights. These capability certificates are short-lived in order to avoid the problems of revocation.

In GT2, the gatekeeper acts as the resource gateway: it allows access only to Grid users who appear in the grid-mapfile. In our current work we make the job manager an AEF as well, by enabling it to enforce policy about fine-grained job access.

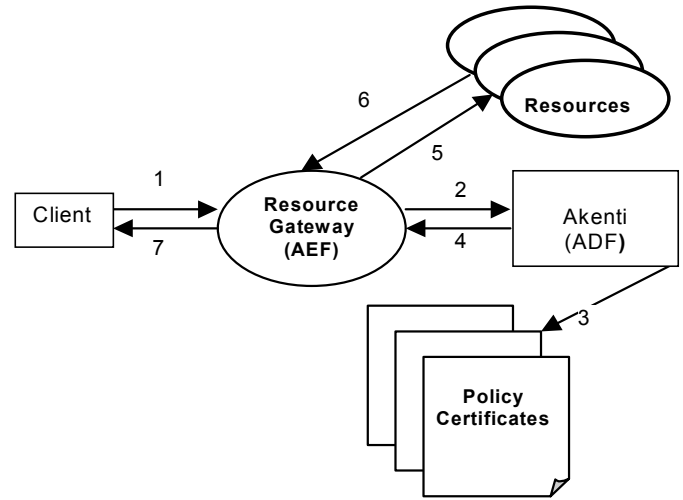

Figure 2 Akenti authorization model in pull mode

### 4.2. Akenti Policy Language

Akenti policy is expressed in XML and stored in three types of signed certificates: *policy certificates, use-condition certificates* and *Akenti attribute certificates* [11]. Policy certificates specify the sources of authority for the resource. Use-condition certificates contain the constraints that control access to a resource. Attribute certificates assign attributes to users that are needed to satisfy the use constraints. The root policy certificate is self-signed and defines the root of trust for the resource domain, so it must be kept in a known secure place. The remainder of the certificates can be stored at distributed sites, since their signatures will be evaluated whenever they are used.

Use-condition certificates contain a Boolean expression specifying what attributes a client must have to be allowed a specific set of actions on the resource. Attributes can be components of the client's DN, including the Common Name (*CN*), which can be used to grant actions to a single individual. They can be *AKENTI* attributes such as role, group and training level or they can be *SYSTEM* attributes such as time of day or load factor on a machine. Thus a constraint might look like the following:

(DN=/O=DOEGrids/OU=People/CN=Jane Doe) ||
(role=developer && (time>5pm) && (time<8am)) ||
(group=clients && executable=TRANSP)
actions=start

This constraint allows Jane Doe to start any job at any time, allows clients who have the role of developer to run any executable between 5 pm and 8 am, and allows members of the client's group to run a specific service, TRANSP, at any time.

The X.509 DN attribute is taken from the client's X.509 certificate. The AKENTI attributes, role and group, are defined by Akenti attribute certificates. Time and executable are SYSTEM attributes and may need to be evaluated by the AEF. In this case, Akenti will return the required attribute value pairs along with the actions that would be allowed if they are satisfied, as conditional actions.

Multiple use-conditions can apply to the same resource. Privileges granted by use-conditions are additive with one major exception. If a use-condition is marked *critical*, a client must satisfy it, or the client will be granted no access, regardless of any other use-conditions.

Policy certificates define the basic trust relationships and are used to bootstrap and to provide closure for the trust chain by specifying the sources of authority for a resource. The sources of authority are the CAs, who are trusted to sign X.509 certificates for all the principals involved in an authorization decision; attribute authorities, who can issue attribute certificates for a user, and the stakeholders, who are allowed to issue use-condition certificates for the resource. Whenever a certificate is used, the Akenti policy engine checks that it has been signed by an acceptable issuer and that the signature verifies. The CAs are represented by their X.509 certificates, which provide a trusted copy of their public keys and information about where they publish certificates and certificate revocation lists. Each stakeholder is represented by a DN and the DN of the CA that issued a certificate for that name, and a list of places, specified by URLs, where the stakeholder puts the use-condition certificates issued. A policy certificate may optionally contain a list of URLs in which to search for attribute certificates.

Authorization policy is associated with individual or collections of resources. Hierarchical resources can inherit policy from parents. Allowing a policy to apply to collections of resources is necessary to scale to more than a handful of resources.

## 5. INTEGRATION OF AKENTI AND JOB MANAGER

In this section we describe how we integrated the Globus Toolkit job manager with Akenti.

### 5.1. Code Integration

While the Globus gatekeeper currently acts as the AEF and ADF for job submission, we decided to add our callout for fine-grained access control to the GT2 job manager [9] for two reasons. First, the job manager is the component that parses the Resource Specification Language (*RSL*) [2] of the job request. RSL consists of attribute value pairs specifying job parameters such as executable description (name, location, etc.), and resource requirements (number of CPUs to be used, maximum allowable memory, etc.). These were the attributes that we wanted to control. Second, the job manager decides and enforces access policy for job control. Requests to terminate, signal or query a job go directly to the job manager via the job handle URL that is returned on job creation. In GT2 the job manager allows these actions only if the requestor has the same Grid id as the job initiator. These were the other actions we wanted to control.

Specifically, our additions consisted of the following:

- Authorization callout API. We designed a callout API to integrate an ADF with the job manager. The callout passes to the ADF module all the information relevant to access control, such as the credential of the user requesting a remote job, the credential of the user who originally started the job, the action to be performed (such as start or cancel a job), a unique job identifier, and the job description expressed in RSL. The ADF responds through the callout API with either success or an appropriate authorization error. This call is made whenever an action needs to be authorized, that is, before instantiating a job and before canceling, querying, or signaling a running job.
- Policy-based authorization for job management. As discussed in Section 3, each job management request other than job start is currently authorized by the job manager so that only the user that started a job is allowed to manage it. We modified the authorization in GRAM to enable Grid users other than the job initiator to manage the job based on policy with decisions rendered through the authorization callout API. In addition to changes to the authorization model, this modification also required extensions to the GRAM client to allow one user to signal a job manager instance owned by another user.
- RSL parameters. We extended RSL to add the "jobtag" parameter allowing the user to submit a job to a specific job management group. If the user does not provide a job tag on start, a default one will be assigned to the job.
- Error reporting. We further extended the GRAM protocol to return authorization errors describing reasons for authorization denial as well as authorization system failures.

In order to provide for easy integration of third-party authorization solutions, the job manager allows callouts to be configurable at run time. Callouts can be configured through either a configuration file or an API call. Configuration consists of specifying an abstract callout name, the path to the dynamic library that implements the callout, and the symbol for the callout in the library. Callouts are invoked through runtime loading of dynamic libraries using GNU Libtool's dlopen-like portability library. Arguments to the callout are passed using the C variable argument list facility. The insertion of callout points into the job manager required defining a GRAM authorization callout type, that is, an abstract callout type, the exact arguments passed to the callout and a set of errors the callout may return. These callout points are configured by parsing a global configuration file.

## 5.2. Authorization Policy

When the job manager calls Akenti, the access decision is based on the Akenti authorization policy. Akenti organizes policy according to the resources that are being controlled. Hence, the first step in writing policy is to determine the set of resources. In the case of fine-grained control of Globus Toolkit job submission, the things that can be controlled are the right to execute a job on a machine, which binaries may be executed, RSL parameters such as requested CPU time, requested scheduling queue, and the rights to stop resume, cancel, or query currently executing jobs.

From the viewpoint of the FusionGrid resource provider, some of these are more important than others and some are hard to enforce:

- Right to submit any job to machine – already enforced by gatekeeper
- Right to start a specific binary – important and can be enforced by the job manager
- Right to limit CPU cycles for a specific job – currently not important, would need to be enforced by the run queue manager (PBS)
- Right to restrict a user or group to a total CPU limit per month – may be important, requires an accounting system
- Right to choose an execution queue – may be important for service guarantees
- Need for at least one class of administrative users who can kill any job – important
- Need for multiple administrative classes that can kill a restricted set of jobs – possibly useful but requires users to understand job classes.

From the Akenti policy point of view these resources can be loosely grouped into machine/site, executables, and jobs. A major consideration in writing a comprehensible policy is to have as little of it as possible. Determining the

```xml
<?xml version="1.0" encoding="US-ASCII"?>
<AkentiCertificate xmlns:xsi="http://www.w3.org/2001/XMLSchema-instance"
       xsi:noNamespaceSchemaLocation='http://www-itg.lbl.gov/Akenti/docs/AkentiCertificate.xsd'>
  <SignablePart>
  <Header Type="Policy" SignatureDigestAlg="RSA-MD5" CanonAlg="Ak1CanAlg" Version="2">
    <UID>"rocky.lbl.gov#104b8965#Thu May 03 17:15:30 PDT 2001"</UID>
    <Issuer>
       <UserDN>/O=doesciencegrid.org/OU=People/CN=Mary R. Thompson</UserDN>
       <CADN>/DC=net/DC=es/OU=Certificate Authorities/OU=DOE Science Grid/CN=pki1</CADN>
     </Issuer>
    <ValidityPeriod Begin="010504001529Z" End="050504001529Z"/>
  </Header>
  <PolicyCert>
    <ResourceName>TRANSP</ResourceName>
    <CAInfo>
      <CADN>/DC=net/DC=es/OU=Certificate Authorities/OU=DOE Science Grid/CN=pki1</CADN>
      <X509Certificate>
         MIICvzCCAiigAwIBAgIBETANBgkqhkiG9w0BAQUFADBbMRkwFwYDVQQKExBET0Ug...
      </X509Certificate>
      <IdDirs> <URL>file:/p/fusiongrid/idCerts</URL></IdDirs>
      <CRLDirs> <URL>ldap://ldap.doegrids.org</URL></CRLDirs>
    </CAInfo>
    <UseCondIssuerGroup>
     <Principal>
        <UserDN>/O=doesciencegrid.org/OU=People/CN=Mary R. Thompson/UserDN>
        <CADN>/DC=net/DC=es/OU=Certificate Authorities/OU=DOE Science Grid/CN=pki1</CADN>
     </Principal>
     <Principal>
        <UserDN>/O=doesciencegrid.org/OU=People/CN=Lew Randerson</UserDN>
        <CADN>/DC=net/DC=es/OU=Certificate Authorities/OU=DOE Science Grid/CN=pki></CADN>
     </Principal>
     <URL>file:/p/fusiongrid/certs</URL>
    </UseCondIssuerGroup>
    <AttrDirs>
       <URL>file:/p/fusiongrid/certs</URL>
    </AttrDirs>
    <CacheTime>3600</CacheTime>
  </PolicyCert>
  </SignablePart>
  <Signature>This is a fake signature</Signature>
</AkentiCertificate>
```

Figure 3 Top-level policy certificate for TRANSP

optimal grouping of resources that can be controlled by a single policy is essential for a concise policy. Since Akenti resources and policies can be hierarchical, the obvious top level is the machine or in the case of a site with several server machines, the site. Policy written for top levels can be inherited by lower levels, so any coarse-grained requirements, for example, the acceptable CAs to issue the client certificates or membership in a VO can be specified there. In the case of the FusionGrid two independent sites are running different codes. One of the sites has two machines dedicated to running its code: a production machine and a more development-oriented machine.

The grouping of executables depends on how many different individual programs are to be run and whether there are obvious classes of programs that can be controlled by a common policy. In the FusionGrid each site supports one main production code. There may also be development versions of the code that should be accessible to a more limited group of users. In addition, users need access to a few simple Unix utilities, such as /bin/date, in order to quickly test that their remote access configuration is working correctly.

Treating jobs as resources is a bit tricky because they are dynamically created objects for which we want to write a static policy. However, it is logical to control jobs based on some characteristic of the job, rather than by specific job instance. Running jobs could be identified by their initiator or by the file that is being executed, or they could be placed into an administrative category when they are started. The last choice lets us write policy about who can control jobs in a given category and gives us the most flexibility over how we want to control jobs. It did require an addition to the original RSL parameters to allow a user to specify a job category when the job was started. The basic Globus Toolkit policy of letting whoever started a job control it requires continued support

### 5.3. Policy for the FusionGrid

The policy we designed to control access to the TRANSP [13] code running at the Princeton Plasma Physics Laboratory has two levels, with several branches at the lower level. There is a *sitewide level* that is named "`TRANSP.`" Policy at this level specifies the CAs that will be trusted to issue X.509 certificates, the stakeholders for the other resources, and the location of the use-condition and attribute certificates. There is also a *subordinate level* that contains separate policies for each class of executables, for example, the production code, test utilities, a development version of the code, and policies for each job category (at the moment we have only one job category). The name of the executable given as an argument to globus-job-run needs to be mapped to an Akenti "resource." We use the following (abbreviated) mapping file to accomplish this:

```
/bin/date   TRANSP/test
/bin/sleep  TRANSP/test
```

```
/p/fusiongrid/trpstart   TRANSP/production
/p/fusiongrid/trspkill   TRANSP/production
/p/fusiongrid/new/trspstart
        TRANPS/development
jobclass  /p/fusiongrid/jobpolicy
```

The complete policy certificate at the top level is shown in Figure 3. It specifies the trusted CAs and where they publish certificates and CRLs, <CAInfo>; the stakeholders and where they publish their use-conditions, <UseCondIssuerGroup>; directories to be searched for attribute certificates, <AttrDirs>; and the maximum caching time for any certificates used in an authorization decision, <CacheTime>. The header of this certificate, and all Akenti certificates, has the type of the certificate, a unique id for the certificate, the issuer who signed the certificate, and a validity period.

Four user groups are granted specific rights: *general* – used for middleware testers, *clients* – physicists who are allowed to run the production code, *developers* – who can run experimental versions of the code, and *administrators* – who can control other users' jobs. Users get the rights of all the groups of which they are members.

Use conditions are written for each class of executables and job category. A portion of a use condition that grants users in the client group to start the production code is shown in Figure 4. Note that the AttributeInfo element includes the authority that is allowed to assert that a user is in the client group.

```
<UseConditionCert critical="false" scope="sub-
tree">
  <ResourceName>TRANSP/production</ResourceName>
  <Condition>
    <Constraint>group = clients</Constraint>
      <AttributeInfo type="AKENTI">
        <AttrName>group</AttrName>
        AttrValue>clients</AttrValue>
        <Principal>
          <UserDN>/O=doesciencegrid.org
            /OU=People/CN=Lew Randerson
          </UserDN>
          <CADN>/DC=net/DC=es/OU=Certificate
            Authorities/OU=DOE Science Grid/
            CN=pki1
          </CADN>
        </Principal>
      </AttributeInfo>
    </Constraint>
  </Condition>
  <Rights>start</Rights>
</UseConditionCert>
```

Figure 4 Use-condition fragment for production code

Figure 5 shows the portion of an attribute certificate that asserts a user's membership in the client group. This certificate had to have been issued and signed by Lew Randerson for it to be accepted by the Akenti policy engine. Note that more than one attribute authority can be specified in a use-condition.

```
<AttributeCert>
   <SubjectAndCA>
     <UserDN>/O=doesciencegrid.org/
       OU=People/CN=Mary R. Thompson
     </UserDN>
     <CADN>/DC=net/DC=es/OU=Certificate
        Authorities/OU=DOE Science Grid
        /CN=pki1
     </CADN>
   </SubjectAndCA>
   <AttrName>group</AttrName>
   <AttrValue>Clients</AttrValue>
</AttributeCert>
```

Figure 5 Attribute certificate fragment

## 6. CONCLUSIONS AND FUTURE WORK

The authorization callout from the GRAM job manager to an Akenti/Globus Toolkit interface module and then to the Akenti authorization server has allowed the FusionGrid to add fine-grained control of the compute services that they are providing. We have experimented with several ways of writing authorization policy and are currently using a scheme based on policy for executables and job classes. So far, the ability of Akenti to support distributed policy created by multiple remote stakeholders has not been used because the code owner and the service provider are the same entity. As a result, all the policy is written by one person and stored in the local file system of the resource host. In the future, NFC members may want to control access to data located at several repositories. In this case there will be two stakeholders for the data, the owner of the repository and the owner of the data, each of whom may want to write policy to control the access to the data. The availability of a GUI to incrementally add to policy by creating a new attribute certificate as new members join the collaboratory has been helpful.

A future goal of the NFC is to provide a high priority service to time critical computations done in support of fusion experiments. One simple way to accomplish this is to write access policy that limits access to the compute resources to a job class that includes only the critical computations. The time period during which the would apply would correspond to the working period of the experiment, typically 8 am to 5 pm. Akenti policy could be written to allow only jobs with the priority class to be run during the these hours and to specify which users are allowed to submit jobs in that class.

## Acknowledgments

We gratefully acknowledge the contributions of Lew Randerson and Doug McCune of the Princeton Physics Plasma Lab in helping to formulate the authorization policy and installing the software at their site. This work was supported by Department of Energy contract with the University of California.DE-AC03-76F00515 and the Mathematical, Information, and Computational Sciences Division subprogram of the Office of Advanced Scientific Computing Research, Office of Science, SciDAC Program, U.S. Department of Energy, under Contract W-31-109-ENG-38. Technical Report number LBNL-52976.